\documentclass[aps,preprint,floatfix,nofootinbib,showpacs]{revtex4-1}

\pdfoutput=1
\usepackage{graphicx,color,float}
\usepackage{hyperref}
\usepackage{multirow}
\usepackage{verbatim}
\usepackage{longtable}
 \usepackage{amsmath}
\usepackage{amsfonts}
\usepackage{amssymb}
\usepackage{rotating}
\usepackage[dvipsnames]{xcolor}

\usepackage{setspace}

\usepackage{ulem}

\newcommand{\imag}{\Im {\rm m}}

\newcommand{\lsim}{\raisebox{-0.13cm}{~\shortstack{$<$ \\[-0.07cm] $\sim$}}~}
\newcommand{\gsim}{\raisebox{-0.13cm}{~\shortstack{$>$ \\[-0.07cm] $\sim$}}~}
\def\slash#1{#1\!\!\!/}

\begin{document}

{\small
\begin{flushright}
IUEP-HEP-22-01
\end{flushright} }

\title{
Impact of the CDF $W$-mass anomaly on two Higgs doublet model
}

\def\slash#1{#1\!\!/}

\renewcommand{\thefootnote}{\arabic{footnote}}

\author{
Yongtae Heo,$^{1,2}$\footnote{yongtae1heo@gmail.com}~
Dong-Won Jung,$^{1,3}$\footnote{dongwon.jung@yonsei.ac.kr}~
and
Jae Sik Lee$^{1,2}$\footnote{jslee@jnu.ac.kr}
}

\affiliation{
$^1$ Department of Physics, Chonnam National University,
Gwangju 61186, Korea\\
$^2$ IUEP, Chonnam National University, Gwangju 61186, Korea \\
$^3$ Department of Physics, 
Yonsei University, Seoul 03722, Korea
}
%\pacs{14.80.Bn.,14.80.Da,14.80.Ec}
%\date{April 12, 2022}
\date{June 6, 2022}
%\date{\color{red}\today}

\begin{abstract}
\begin{spacing}{1.30}
We consider the implication of
the recent CDF $W$-mass anomaly in the framework of 
two Higgs doublet model.
We find that the large deviation of the $S$ and
$T$ parameters from their SM values of zero 
leads to the upper limit of about $1$ TeV on the heavy 
charged and neutral Higgs bosons when it is
combined with
the theoretical constraints from the perturbative unitarity and
for the Higgs potential to be bounded from below.
\end{spacing}
\end{abstract}

\maketitle

\section{Introduction}
Recently, using $8.8$ fb$^{-1}$
data collected at the Fermilab Tevatron collider
with a center-of-mass energy of 1.96 TeV,
the CDF collaboration has reported the result of
the $W$ boson mass measurement with unprecedented precision 
\cite{CDF:2022hxs}
\begin{equation}
80,433.5 \ \pm \ 9.4 \ {\rm MeV}\,,
\end{equation}
which, comparing with the SM expectation of
$80,357 \ \pm \ 6 \ {\rm MeV}$~\cite{ParticleDataGroup:2020ssz},
yields a difference with a significance of $7.0\sigma$.
Performing a global fit of electroweak data with
the high-precision CDF measurement while fixing $U=0$, one may find
the large central values 
of the oblique parameters $S$ and $T$
together with the standard deviations such as
\cite{Lu:2022bgw}
\begin{equation}
\label{eq:STCDF}
(\widehat S_0\,,\ \sigma_S)\ =\ (0.15\,,\ 0.08)\;,\qquad
(\widehat T_0\,,\ \sigma_T)\ =\ (0.27 \,,\ 0.06)\; ,
\end{equation}
and a strong correlation $\rho_{ST}=0.93$ between them.
We find other estimations such as
$S=0.064\pm 0.090$ and $T=0.14\pm 0.064$~\cite{Strumia:2022qkt},
$S=0.100\pm 0.074$ and $T=0.177\pm 0.071$~\cite{deBlas:2022hdk},
$S=0.086\pm 0.076$ and $T=0.167\pm 0.059$~\cite{Paul:2022dds},
$S=0.17$ and $T=0.27$~\cite{Asadi:2022xiy},
$T=0.110\pm 0.018$~\cite{DiLuzio:2022xns}, and
$T:\{0.15949\,, 0.210177\}$~\cite{Babu:2022pdn}.
For further works involved with the $W$-mass anomaly, we refer to
Refs.~\cite{Fan:2022dck,Zhu:2022tpr,Athron:2022qpo,Yang:2022gvz,
Du:2022pbp,Tang:2022pxh,Cacciapaglia:2022xih,Blennow:2022yfm,
Arias-Aragon:2022ats,Sakurai:2022hwh,Fan:2022yly,Liu:2022jdq,
Lee:2022nqz,Cheng:2022jyi,Song:2022xts,Bagnaschi:2022whn,Bahl:2022xzi,
Athron:2022isz,Gu:2022htv,Heckman:2022the}.

In this Letter, 
taking the framework of two Higgs doublet model (2HDM),
we report that the large deviation of the
$T$ parameter, especially, from its SM value of zero
results in the upper limit of about $1$ TeV on the masses of the heavy 
charged and neutral Higgs bosons
when it is combined with 
the theoretical constraints from the perturbative unitarity and
for the Higgs potential to be bounded from below.

\section{Framework}
The general 2HDM scalar potential in the so-called Higgs basis
\cite{Donoghue:1978cj,Georgi:1978ri}
where only one doublet contains the non-vanishing vacuum expectation value
$v$ is given by
\begin{eqnarray}
\label{eq:VHiggs}
V_{\cal H} &=&
Y_1 ({\cal H}_1^{\dagger} {\cal H}_1)
+Y_2 ({\cal H}_2^{\dagger} {\cal H}_2)
+Y_3 ({\cal H}_1^{\dagger} {\cal H}_2)
+Y_3^{*}({\cal H}_2^{\dagger} {\cal H}_1) \nonumber \\
&&+ Z_1 ({\cal H}_1^{\dagger} {\cal H}_1)^2 + Z_2
({\cal H}_2^{\dagger} {\cal H}_2)^2 + Z_3 ({\cal H}_1^{\dagger}
{\cal H}_1)({\cal H}_2^{\dagger} {\cal H}_2) + Z_4 ({\cal H}_1^{\dagger}
{\cal H}_2)({\cal H}_2^{\dagger} {\cal H}_1) \nonumber \\
&&+ Z_5 ({\cal H}_1^{\dagger} {\cal H}_2)^2 +
Z_5^{*} ({\cal H}_2^{\dagger} {\cal H}_1)^2 + Z_6
({\cal H}_1^{\dagger} {\cal H}_1) ({\cal H}_1^{\dagger} {\cal H}_2) + Z_6^{*}
({\cal H}_1^{\dagger} {\cal H}_1)({\cal H}_2^{\dagger} {\cal H}_1) \nonumber \\
&& + Z_7 ({\cal H}_2^{\dagger} {\cal H}_2) ({\cal H}_1^{\dagger} {\cal H}_2) +
Z_7^{*} ({\cal H}_2^{\dagger} {\cal H}_2) ({\cal H}_2^{\dagger} {\cal H}_1)\;,
\end{eqnarray}
which contains 3 dimensionful quadratic and 7 dimensionless quartic parameters of which 
four parameters are complex. 
In this work we consider the CP-conserving case assuming
$\imag(Y_3)=\imag(Z_{5,6,7})=0$
but without imposing the so-called 
${\cal Z}_2$ symmetry.
\footnote{
The ${\cal Z}_2$ symmetry in the Higgs basis
might be realized by requiring the invariance of the Higgs potential 
under the transformations 
${\cal H}_1 \to -{\cal H}_1$ and ${\cal H}_2 \to +{\cal H}_2$.
Therefore, by imposing it strictly,
the $Y_3$, $Z_6$ and $Z_7$ terms have to be removed from 
the Higgs potential.
Without imposing the ${\cal Z}_2$ symmetry,
there appear
the Higgs-mediated flavor-changing neutral currents 
at tree level which
could be avoided, for example,
by considering the models in which
the Yukawa matrices describing the couplings of the
two Higgs doublets to the SM fermions are aligned in the flavor space
\cite{Manohar:2006ga,Pich:2009sp,Penuelas:2017ikk}.}
The complex SU(2)$_L$ doublets of ${\cal H}_1$ and ${\cal H}_2$
can be parameterized as
\begin{eqnarray}
\label{eq:H12inHiggBasis}
{\cal H}_1=
\left(\begin{array}{c}
G^+ \\ \frac{1}{\sqrt{2}}\,(v+\varphi_1+iG^0)
\end{array}\right)\,; \ \ \
{\cal H}_2=
\left(\begin{array}{c}
H^+ \\ \frac{1}{\sqrt{2}}\,(\varphi_2+iA)
\end{array}\right)\,,
\end{eqnarray}
where $v = \left(\sqrt{2}G_F\right)^{-1/2} \simeq 246.22$ GeV and
$G^{\pm,0}$ and $H^\pm$ stand for the
Goldstone and charged Higgs bosons, respectively.
For the neutral Higgs bosons, 
$A$ denotes a CP-odd mass eigenstate and
the two states
$\varphi_1$ and $\varphi_2$ result in
two CP-even mass eigenstates through mixing and one of them should
play the role of the SM Higgs boson.
The tadpole conditions relate the quadratic parameters $Y_{1,3}$
to  $Z_{1,6}$ as follows:
\begin{equation}
\label{eq:higgsbasistadpole}
Y_1 \ + \ Z_1 v^2\ = 0 \,; \ \ \
Y_3 \ + \ \frac{1}{2}Z_6 v^2\ = 0 \,.
\end{equation}
The 2HDM Higgs potential includes the mass
terms which can be cast into the form consisting of three parts
\begin{equation}
V_{{\cal H}\,, {\rm mass}}=
M_{H^\pm}^2 H^+ H^- \ + \ 
\frac{1}{2}\,M_A^2\, A^2 \ + \ 
\frac{1}{2}
(\varphi_1 \ \varphi_2)\,{\cal M}^2_0\,
\left(\begin{array}{c}
\varphi_1 \\ \varphi_2  \end{array}\right)\,,
\end{equation}
in terms of the charged Higgs bosons $H^\pm$,
the neutral CP-odd Higgs boson $A$, and the two
neutral CP-even scalars $\varphi_{1,2}$.
The charged and CP-odd Higgs boson masses are given by
\begin{equation}
M_{H^\pm}^2= Y_2 +\frac{1}{2} Z_3 v^2\,, \ \ \
M_A^2=M_{H^\pm}^2+ \left(\frac{1}{2}Z_4 -Z_5\right)v^2\,,
\end{equation}
while the $2\times 2$ mass-squared matrix of the neutral Higgs
bosons ${\cal M}_0^2$ takes the form
\begin{equation}
{\cal M}^2_0 = \left(\begin{array}{cc}0  & 0 \\ 0 & M_A^2
\end{array}\right)
\ + \
\left(\begin{array}{cc}
2 Z_1 & Z_6  \\
Z_6 & 2 Z_5 
\end{array}\right)\,v^2\,.
\end{equation}
Note that the quartic couplings $Z_2$ and $Z_7$ have nothing to
do with the masses of Higgs bosons and the mixing of the neutral ones.
We further note that
$\varphi_1$ does not mix with
$\varphi_2$ in the $Z_6=0$ limit,
and its mass squared is simply given by $2 Z_1 v^2$ which
%gives $Z_1\simeq 0.13\,(M_{h_{\rm SM}}/125.5\,{\rm GeV})^2$.
gives $Z_1\simeq 0.13$.

With the $2\times 2$ real and symmetric mass-squared
${\cal M}_0^2$ is given, the mixing is described by
\begin{eqnarray}
(\varphi_1,\varphi_2)^T_\alpha=O_{\alpha i} (h,H)^T_i\,,
\end{eqnarray}
such that $O^T {\cal M}_0^2 O={\rm diag}(M_h^2,M_H^2)$ with
the orthogonal mixing matrix $O$ parameterized as
\begin{equation}
O \ = \ \left(\begin{array}{cc}
c_\gamma & s_\gamma  \\
-s_\gamma & c_\gamma 
\end{array}\right)\,,
\end{equation}
introducing the mixing angle $\gamma$ between the
two CP-even states $\varphi_1$ and $\varphi_2$.
Then the quartic couplings $\{Z_1,Z_4,Z_5,Z_6\}$ are given by
\begin{eqnarray}
\label{eq:Z1456_cpc}
Z_1 &=& \frac{1}{2v^2}\left(c_\gamma^2 M_h^2 + s_\gamma^2 M_H^2 \right)\,, \ \ \
\hspace{1.1cm}
Z_4  =  \frac{1}{v^2}\left( s_\gamma^2 M_h^2 + c_\gamma^2 M_H^2
+M_A^2 -2M_{H^\pm}^2 \right)\,,\nonumber \\[2mm]
Z_5 &=& \frac{1}{2v^2}\left(s_\gamma^2 M_h^2
+c_\gamma^2 M_H^2 -M_A^2 \right)\,, \ \ \
Z_6  =  \frac{1}{v^2}\left(-M_h^2 + M_H^2 \right)c_\gamma s_\gamma\,,
\end{eqnarray}
in terms of the four masses  $M_{h,H,A,H^\pm}$ and the mixing angle $\gamma$.
We observe that, in the alignment limit of $\sin\gamma=0$,
$Z_1=M_h^2/2v^2$ and $Z_6=0$
and $Z_4$ and $Z_5$ are determined by the mass differences of
$M_H^2+M_A^2-2M_{H^\pm}^2$ and $M_H^2-M_A^2$, respectively
\cite{Gunion:2002zf,Craig:2013hca,Carena:2013ooa,Dev:2014yca,Bernon:2015qea}.
For the study of the CP-conserving case,
one may choose one of the following two equivalent sets:
\begin{eqnarray}
\label{eq:input_cpc}
{\cal I}&=&
\left\{v,Y_2;M_{H^\pm},M_h,M_H,M_A,\gamma ;Z_2,Z_7\right\}\,,
\nonumber \\[2mm]
{\cal I}^\prime&=&
\left\{v;M_{H^\pm},M_h,M_H,M_A,\gamma ;Z_3;Z_2,Z_7\right\}\,,
\end{eqnarray}
each of which contains 9 real degrees of freedom,
and the convention of $|\gamma|\leq \pi/2$ can be taken
without loss of generality resulting in
$c_\gamma \geq 0$ and ${\rm sign}(s_\gamma)={\rm sign}(Z_6)$
assuming $M_H>M_h=125.5$ GeV.
The heavy Higgs masses squared are scanned up to $(1.5~{\rm TeV})^2$
and the quartic couplings $Z_{2}$, $|Z_{3}|$, and $|Z_{7}|$ 
up to 3, 10, 5, respectively.

\section{Analysis}
First, we consider the  perturbative unitarity (UNIT) conditions and
those for the Higgs potential to be bounded from below (BFB) to obtain the
primary theoretical constraints on the potential parameters 
or, equivalently, the constraints on the Higgs-boson masses including 
correlations among them and the mixing angle $\gamma$.
For the unitarity conditions, we closely follow Ref.~\cite{Jurciukonis:2018skr}
taking into account three scattering matrices
which are expressed in terms of the quartic couplings $Z_{1-7}$.
Using the set ${\cal I}^\prime$ in Eq.~(\ref{eq:input_cpc})
for the input parameters,
all the seven quartic couplings are fixed exploiting the relations
given by Eq.~(\ref{eq:Z1456_cpc}). 
For the details of the implementation of the UNIT conditions, we refer to 
Ref.~\cite{Choi:2021sot}.
For the BFB constraints, we require the following 5 necessary conditions
for the Higgs potential to be bounded-from-below 
\cite{Branco:2011iw}:
\begin{eqnarray}
\label{eq:bfb}
Z_1 \geq 0 \,, \ \ \ Z_2 \geq 0\,;&& \nonumber \\[2mm]
2\sqrt{Z_1Z_2}+Z_3 \geq  0 \,, \ \ \  2\sqrt{Z_1Z_2}+Z_3+Z_4-2|Z_5|\geq 0\,;&& \nonumber \\[2mm]
Z_1+Z_2+Z_3+Z_4+2|Z_5|-2|Z_6+Z_7|  \geq 0\,.&&
\end{eqnarray}

Second, we consider
the electroweak (ELW) oblique corrections to the so-called $S$, $T$ and $U$
parameters~\cite{Peskin:1990zt,Peskin:1991sw} which provide significant
constraints on the quartic couplings of the 2HDM.
Fixing $U=0$ which is suppressed by
an additional factor $M_Z^2/M^2_{\rm BSM}$
\footnote{Here, $M_{\rm BSM}$ denotes some heavy mass scale involved
with new physics beyond the Standard Model.} relative to 
$S$ and $T$,
the $S$ and $T$ parameters are constrained as follows
\cite{ParticleDataGroup:2020ssz,Lee:2012jn}
\begin{equation}
\label{eq:STRange}
\frac{(S-\widehat S_0)^2}{\sigma_S^2}\ +\
\frac{(T-\widehat T_0)^2}{\sigma_T^2}\ -\
2\rho_{ST}\frac{(S-\widehat S_0)(T-\widehat T_0)}{\sigma_S \sigma_T}\
\leq\ R^2\,(1-\rho_{ST}^2)\; ,
\end{equation}
with $R^2=2.3$, $4,61$, $5.99$, $9.21$, $11.83$ at $68.3 \%$, $90 \%$,
$95 \%$, $99 \%$, and $99.7 \%$  confidence levels (CLs), respectively.
For our numerical analysis, we take the 95\% CL limit. 
For the central values $\widehat S_0$ and $\widehat T_0$ and
the standard deviations $\sigma_{S,T}$, we adopt those
given in Ref.~\cite{Lu:2022bgw}, see Eq.~(\ref{eq:STCDF}).

Using the set ${\cal I}^\prime$ for the input parameters,
the $S$ and $T$ parameters 
take the following forms
\cite{Toussaint:1978zm,Grimus:2007if,Grimus:2008nb,Branco:2011iw}:
\begin{eqnarray}
\label{eq:STCPC}
S \!&=&\! -\frac{1}{4\pi} \left[
F^\prime_\Delta(M_{H^\pm},M_{H^\pm})
-c_\gamma^2\,F^\prime_\Delta(M_A,M_H)
-s_\gamma^2\,F^\prime_\Delta(M_A,M_h)
\right]\,, \nonumber\\
T \!&=&\! \frac{\sqrt{2}G_F}{16\pi^2\alpha_{\rm EM}}\ [
F_\Delta(M_A,M_{H^\pm})+c_\gamma^2\,F_\Delta(M_H,M_{H^\pm})+
s_\gamma^2\,F_\Delta(M_h,M_{H^\pm}) \nonumber \\
&& \hspace{4.6cm}
-c_\gamma^2\,F_\Delta(M_A,M_H) -s_\gamma^2\,F_\Delta(M_A,M_h)
]\; .\quad
\end{eqnarray}
The one-loop functions are given by
\footnote{See, for example, Ref.\cite{Kanemura:2011sj}.}
\begin{eqnarray}
\label{eq:ffp}
F_\Delta(m_0,m_1) &=& F_\Delta(m_1,m_0) =
\frac{m_0^2+m_1^2}2 -\frac{m_0^2m_1^2}{m_0^2-m_1^2}\ln\frac{m_0^2}{m_1^2}\,,
\nonumber \\[3mm]
F_\Delta^\prime(m_0,m_1) &=& F_\Delta^\prime(m_1,m_0) =
-\frac{1}{3} \left[ \frac{4}{3}
-\frac{m_0^2 \ln m_0^2 -m_1^2 \ln m_1^2}{m_0^2-m_1^2}
-\frac{m_0^2+m_1^2}{(m_0^2-m_1^2)^2}F_\Delta(m_0,m_1) \right]\,.
\end{eqnarray}
With $F_\Delta(m,m)=0$ and $F_\Delta^\prime(m,m)=\frac{1}{3}\ln m^2$,
we observe that 
$T$ is identically vanishing when $M_{H^\pm}=M_A$.

\begin{figure}[t!]
\vspace{-1.0cm}
\begin{center}
\includegraphics[width=8.5cm]{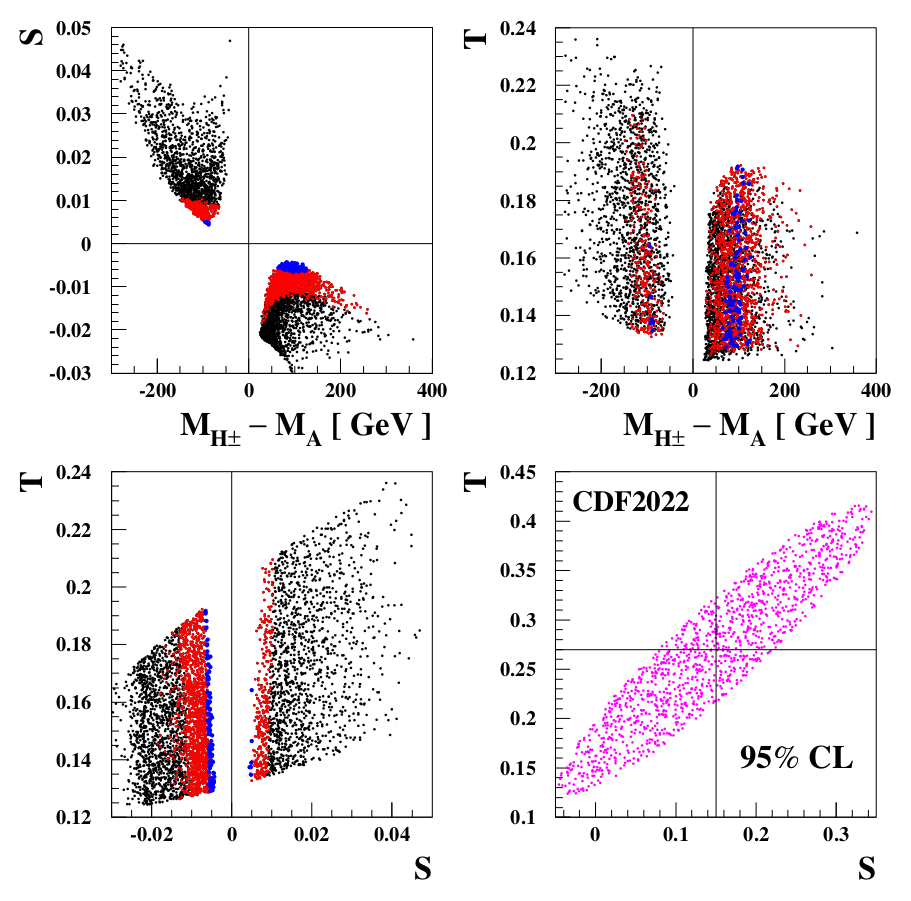}
\includegraphics[width=8.5cm]{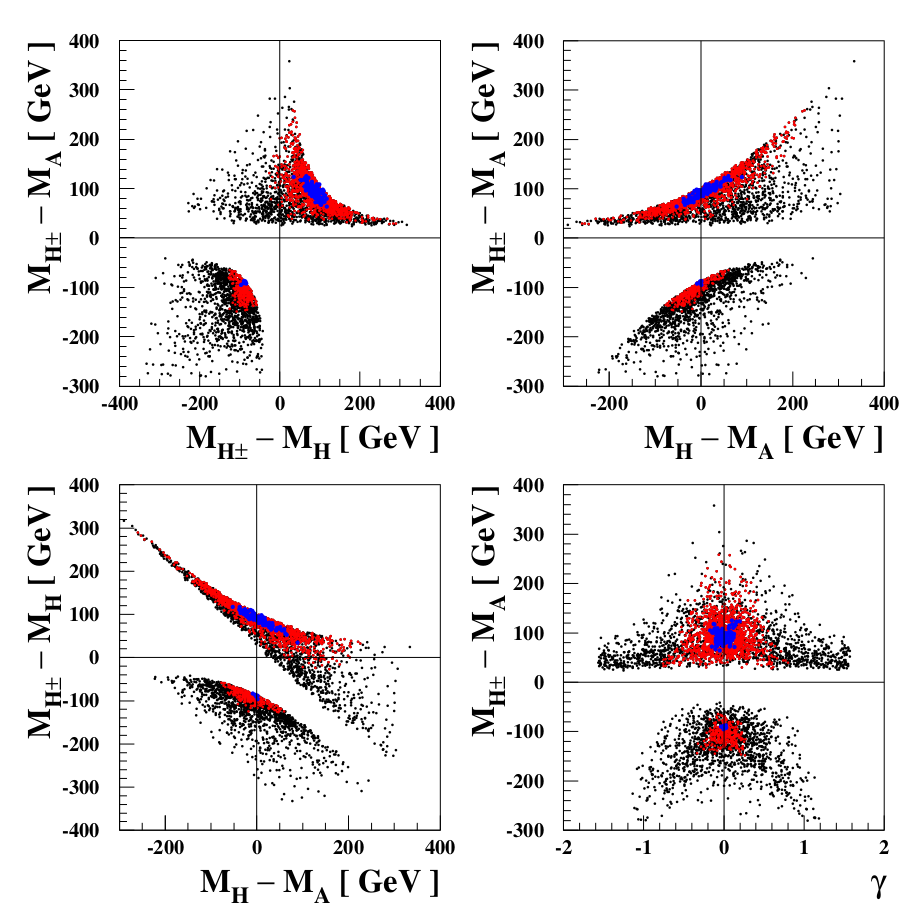}
\end{center}
\vspace{-0.5cm}
\caption{\it
Correlations among the $S$ and $T$ parameters, mass differences, 
and the mixing
angle $\gamma$ using the set ${\cal I}^\prime$ imposing
the combined UNIT$\oplus$BFB$\oplus$ELW$_{95\%}$ constraints.
(Left) 
Scatter plots of
$S$ versus $M_{H^\pm}-M_A$ (upper left),
$T$ versus $M_{H^\pm}-M_A$ (upper right), and
$T$ versus $S$ (lower left).
The red and blue points are for
$M_{H^\pm}>500$ GeV and $M_{H^\pm}>900$ GeV, respectively.
In the lower-right plot, as a reference,
the 95\% CL ELW constraint on the $S$ and $T$
parameters according to Eqs.~(\ref{eq:STRange}) and (\ref{eq:STCDF}) 
is shown.
(Right) Scatter plots of
$M_{H^\pm}-M_A$ versus $M_{H^\pm}-M_H$ (upper left),
$M_{H^\pm}-M_A$ versus $M_{H}-M_A$ (upper right),
$M_{H^\pm}-M_H$ versus $M_{H}-M_A$ (lower left), and
$M_{H^\pm}-M_A$ versus $\gamma$ (lower right).
The red and blue points are again for
$M_{H^\pm}>500$ GeV and $M_{H^\pm}>900$ GeV, respectively.}
\label{fig:stmm}
\end{figure}
In the left panel of Fig.~\ref{fig:stmm}, we show the $S$ and $T$
parameters imposing the UNIT, BFB, and ELW constraints abbreviated by
the combined UNIT$\oplus$BFB$\oplus$ELW$_{95\%}$ ones.
Note that the 95\% CL ELW limits are adopted and
the heavy Higgs masses squared are scanned up to $(1.5~{\rm TeV})^2$.
We find that $S$ takes values in the range
between $-0.03$ and $0.05$ whose
absolute values are smaller than $\sigma_S=0.08$, see Eq.~(\ref{eq:STCDF}).
Note that the narrow region around $0$ with radius about $0.004$ 
is not allowed since the misalignment between $M_{H^\pm}$ and $M_A$
is required to achieve the sizable central value of the $T$ parameter,
see the first line of Eq.~(\ref{eq:STCPC}).
Note that $S$ is negative (positive) when $M_{H^\pm} > (<) M_A$.
The $T$ parameter takes its value between $0.12$ and $0.24$.
Note that $T$ is positive definite and  sizable and, accordingly,
$M_{H^\pm}=M_A$ is not allowed. Actually, we find that
the region $-40 \lsim (M_{H^\pm}-M_A)/{\rm GeV} \lsim 20$ 
is ruled out at 95\% CL.
In the right panel of Fig.~\ref{fig:stmm}, we show the
correlations among the mass differences and the mixing
angle $\gamma$. As $M_{H^\pm}$ increases,
the mass difference between the charged and neutral Higgs bosons
$|M_{H^\pm}-M_{A,H}|$ converges to the value of about 100 GeV.
On the other hand, we find that 
$|M_H-M_A|/{\rm GeV}\lsim 250\,(50)$ GeV when
$M_{H^\pm}>500\,(900)$ GeV.

\begin{figure}[t!]
\vspace{-1.0cm}
\begin{center}
\includegraphics[width=8.5cm]{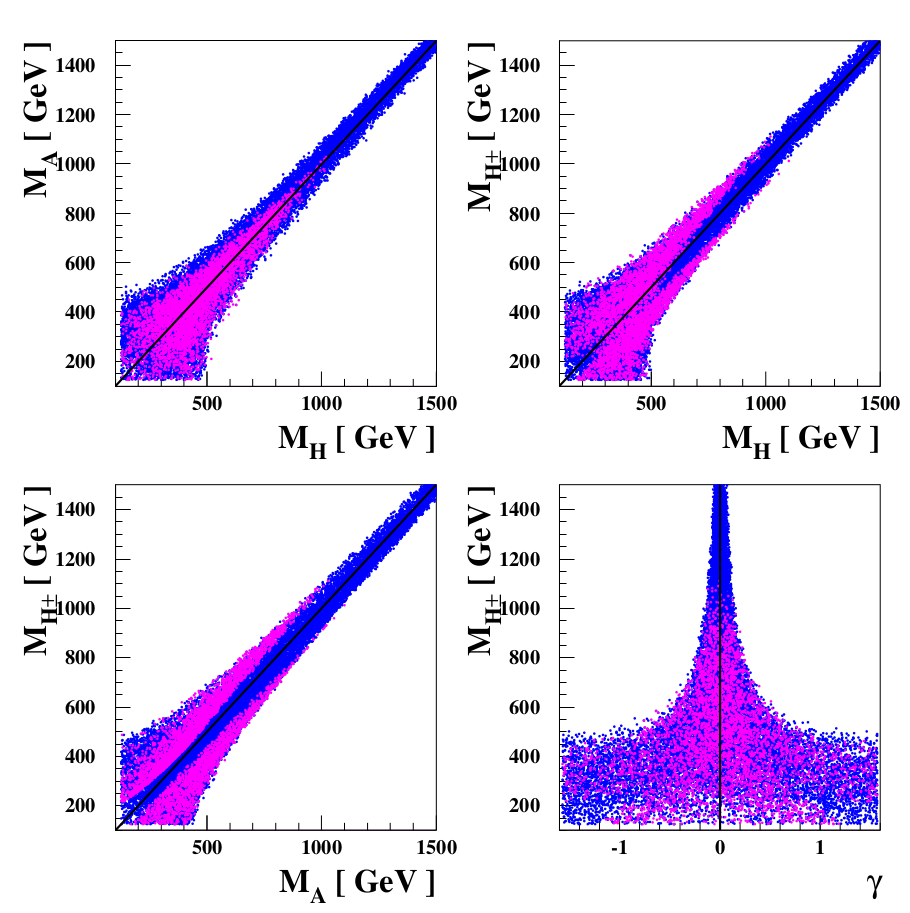}
\includegraphics[width=8.5cm]{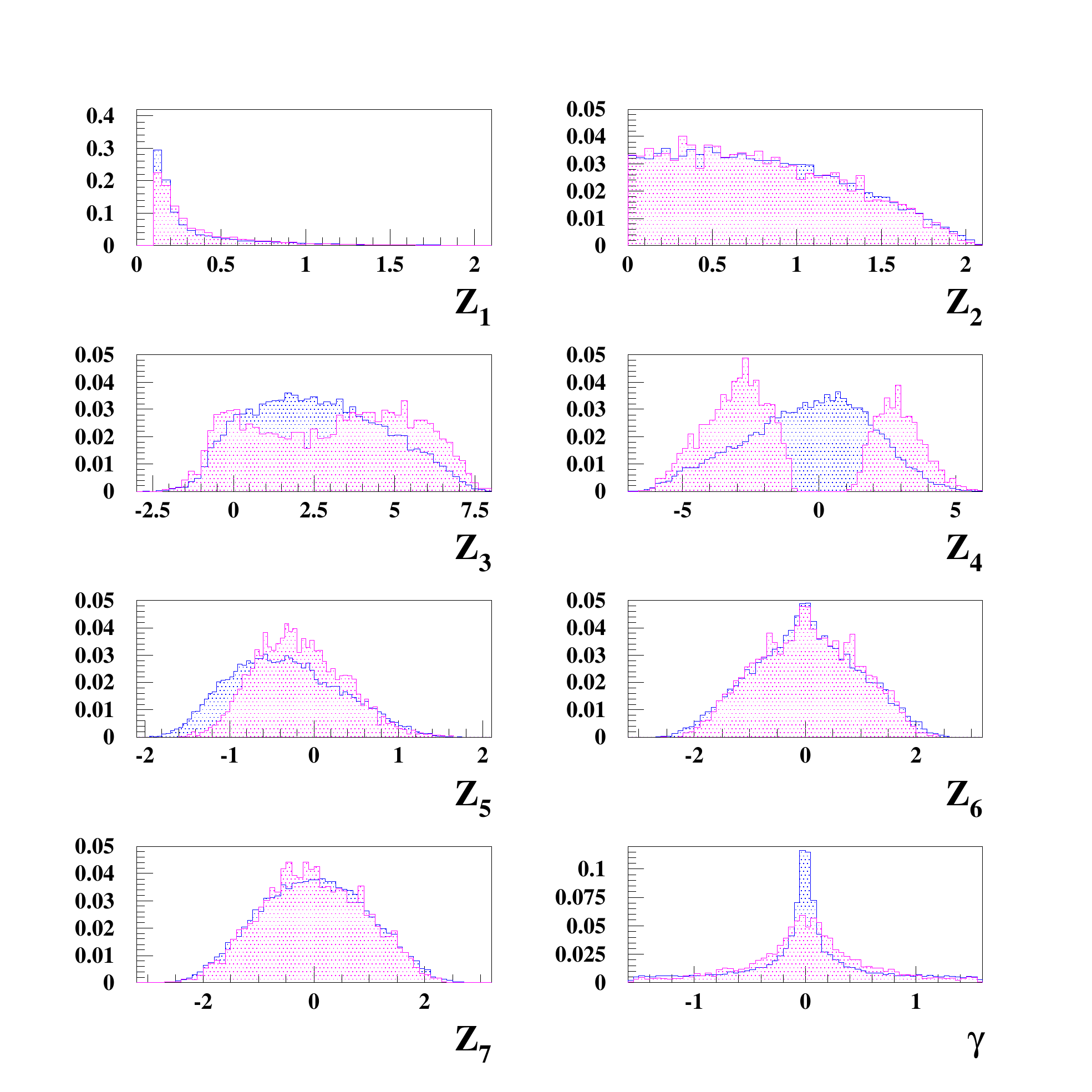}
\end{center}
\vspace{-0.5cm}
\caption{\it The UNIT$\oplus$BFB$\oplus$ELW$_{95\%}$ constraints (magenta)
using ${\cal I}^\prime$, see Eq.~(\ref{eq:input_cpc}).
For comparisons, we also show the results after applying
only the UNIT$\oplus$BFB constraints (blue).:
(Left) Scatter plots of $M_A$ versus $M_H$ (upper left),
$M_{H^\pm}$ versus $M_H$ (upper right),
$M_{H^\pm}$ versus $M_A$ (lower left), and
$M_{H^\pm}$ versus $\gamma$ (lower right).
%with the UNIT$\oplus$BFB (blue) and
%UNIT$\oplus$BFB$\oplus$ELW$_{95\%}$ (magenta) constraints imposed.
%
(Right) The normalized distributions of the quartic couplings
and the mixing angle $\gamma$.
%obtained by considering
%all the three constraints of
%UNIT$\oplus$BFB$\oplus$ELW$_{95\%}$ (magenta).
}
\label{fig:mmziang}
\end{figure}
We show the correlations among the heavy Higgs-boson masses and the mixing
angle $\gamma$ in the left panel of Fig.~\ref{fig:mmziang}.
Requiring the ELW constraint in addition to the UNIT$\oplus$BFB ones,
we find that $Z_1$ and $\gamma$ take values near to 0 
less likely while 
$Z_5$ positive ones more likely,
see the right panel of Fig.~\ref{fig:mmziang}.
On the other hand, the
$Z_2$ and $Z_7$ distributions remain almost the same since they
are irrelevant  to the masses of Higgs bosons and the mixing 
angle $\gamma$.
The $Z_3$ and $Z_6$ distributions undergo some changes but
the $Z_4$ distribution changes most drastically 
excluding the region $|Z_4|\lsim 1$.
This could be understood by looking into the expression for
$Z_4$ given in Eq.~(\ref{eq:Z1456_cpc}).
Taking $\gamma=0$ and $M_H=M_A$ for the convenience of discussion, 
one may have
\begin{equation}
Z_4 v^2 
%=(M_H-M_{H^\pm}) (M_H+M_{H^\pm}) +
%(M_A-M_{H^\pm}) (M_A+M_{H^\pm})  
=4 \Delta_M\, \overline{M}\,,
\end{equation}
with $\Delta_M\equiv M_A-M_{H^\pm}$ and
$\overline M\equiv (M_A+M_{H^\pm})/2$. Therefore $Z_4$ can not vanish 
in the presence of
the misalignment between $M_{H^\pm}$ and $M_A$, which is
required to achieve the sizable central value of the $T$ parameter.
We recall that the mass difference $|\Delta_M|$ approaches to $100$ GeV
as $M_{H^\pm}$ grows, see the right panel of Fig.~\ref{fig:stmm}.
On the other hand, the above relation could be rewritten for $\overline M$
as
\begin{equation}
\overline{M} \ = \ \frac{Z_4\,v^2}{4\,\Delta_M}
\ \leq \ \frac{v^2}{4}\,
\left(\frac{Z_4}{\Delta_M}\right)_{\rm max}
\ \leq \ \frac{v^2}{4}\,
\frac{|Z_4|_{\rm max}}{|\Delta_M|_{\rm min}}\,,
\end{equation}
which implies that there {\it exists} the absolute
upper limit on the masses of the 
heavy charged and neutral Higgs bosons with
$|Z_4|_{\rm max}$ and $|\Delta_M|_{\rm min}$
from the UNIT$\oplus$BFB and ELW$_{95\%}$ constraints,
respectively.
We observe that $|Z_4|$ tends to increase as $M_{H^\pm}$ grows
until it reaches $\sim 6$ where $M_{H^\pm}$ takes 
its maximum value of about 1 TeV.
When $M_{H^\pm}$ approaches to 1 TeV,
$|\Delta_M|$ converges to 100 GeV  while
taking its minimum value of about 30 GeV around $M_{H^\pm} = 450$ GeV,
see the upper plots
in the right panel of Fig.~\ref{fig:stmm} and
the plot of $M_{H^\pm}$ versus $M_A$ in the left panel of
Fig.~\ref{fig:mmziang}. 
\footnote{
Note that the precise value of $|\Delta_M|$ 
and its parametric dependence
depend on the order to which the
$S$ and $T$ parameters are computed.
In Ref.~\cite{Lu:2022bgw} from which we are adopting 
the central values and the standard deviations of
the $S$ and $T$ parameters, 
the electroweak oblique parameters are computed at the one-loop order
as in this work.
In Ref.~\cite{Bahl:2022xzi}, a two-loop calculation of the $W$-boson mass
has been performed and it is found that $|\Delta_M|\gsim 50$ GeV.}
Taking into account the full correlations among the masses of
heavy Higgs bosons and the mixing angle, we find
that $(Z_4/\Delta_M)_{\rm max} \sim 6/(100\,{\rm GeV})$ leading to
the upper limit of about 1 TeV as clearly shown
in the left panel of Fig.~\ref{fig:mmziang}.

Finally, in order to assess the reliability of our main result, 
we consider the variation of the upper limit 
on the heavy Higgs-boson masses
by shifting the central value of
the $T$  parameter by the amount of $\pm\sigma_T$.
We find that $M_{H^\pm} \lsim 1,000^{-100}_{+400} $ GeV
taking $\widehat{T}_0=0.27\pm 0.06$.
We observe that the upper limit is quite stable 
when $\widehat{T}_0$ is larger than the nominal value of $0.27$ while
it grows faster for the smaller values of $\widehat{T}_0$.

\section{Conclusions}
We consider the implication of
the recent CDF $W$-mass anomaly in the framework of 2HDM.
We find that the large deviation of the $S$ and
$T$ parameters from their SM values of zero 
leads to the upper limit of about $1$ TeV on the masses of the heavy 
charged and neutral Higgs bosons when it is
combined with
the theoretical constraints from the perturbative unitarity and
for the Higgs potential to be bounded from below.

\bigskip

\noindent
{\it Note added}: After the completion of our work, we 
have received Ref.~\cite{Lee:2022gyf} in which the CDF
$W$-mass anomaly studied in the framework of 2HDMs.
They also find the upper bounds of about 1 TeV
on the masses of the heavy Higgs bosons
by including phenomenological constraints from
flavor observables, Higgs precision data, and 
direct collider search limits  in addition to
theoretical ones.
We observe that the upper limit of about 1 TeV  on the 
masses of heavy Higgs bosons are 
largely unaffected by the phenomenological constraints.

%
%\newpage
%
\section*{Acknowledgment}
This work was supported by the National Research Foundation (NRF) of Korea
Grant No. NRF-2021R1A2B5B02087078 (Y.H., D.-W.J., J.S.L.).
The work of D.-W.J. was also supported in part by
the NRF of Korea Grant Nos. NRF-2019R1A2C1089334 and 
NRF-2021R1A2C2011003 and in part by the Yonsei University Research Fund
of 2022.
The work of J.S.L. was also supported in part by
the NRF of Korea Grant No. NRF-2022R1A5A1030700.

%\newpage
%%%%%%%%%%%%%%%%%%%%%%%%%%%%%%%%%%%%%%%%

\end{document}